# An Exact Path-Loss Density Model for Mobiles in a Cellular System

Mouhamed Abdulla and Yousef R. Shayan
Department of Electrical and Computer Engineering
Concordia University
Montréal, Québec, Canada
Email: {m_abdull, yshayan}@ece.concordia.ca

## ABSTRACT

In trying to emulate the spatial position of wireless nodes for purpose of analysis, we rely on stochastic simulation. And, it is customary, for mobile systems, to consider a base-station radiation coverage by an ideal cell shape. For cellular analysis, a hexagon contour is always preferred mainly because of its tessellating nature. Despite this fact, largely due to its intrinsic simplicity, in literature only random dispersion model for a circular shape is known. However, if considered, this will result an unfair nodes density specifically at the edges of non-circular contours. As a result, in this paper, we showed the exact random number generation technique required for nodes scattering inside a hexagon. Next, motivated from a system channel perspective, we argued the need for the exhaustive random mobile dropping process, and hence derived a generic close-form expression for the path-loss distribution density between a base-station and a mobile. Last, simulation was used to reaffirm the validity of the theoretical analysis using values from the new IEEE 802.20 standard.

## 1. INTRODUCTION

The cellular concept for mobile systems started over 30 years ago, and its importance is even more relevant as we move forward toward 4G systems with WiMAX, LTE, and the new IEEE 802.20 Mobile Broadband Wireless Access (MBWA) [1]. In this contribution, the focus will be on the spatial position of Mobile Stations (MS). And this is an essential parameter in studying a constellation of nodes, because the location of transceivers will directly affect important communication factors such as: network capacity, coverage area, connectivity of terminals, power consumption, and interference, among others.

A cost-effective way to achieve this investigation would be to use random simulation to scatter nodes inside a cell. During, preliminary analysis and design, it is common to consider the electromagnetic emission from an omni-directional Base-Station (BS) antenna to have an ideal geometry such as a circle or a hexagon. In fact, the hexagon shape is more preferred because of its tessellating feature [2]. I spite of this, only a model for mobile dispersion inside a circle is available, as say in [3] and [4]. Therefore, in this paper, we will show and give expressions for modeling exact Random Number Generator (RNG) for stochastic nodes spreading inside a non-sectored and sectored hexagon cell.

Moreover, during system analysis, wireless channel corruption such as Path-Loss (PL) is always vital and critical. Thus, being able to predict the PL behavior through the use of a Probability Density Function (PDF) becomes very useful. However, by and large, the only way to obtain an estimate of the PL density relays on stochastic computationally complex Monte Carlo simulation for each system being researched. As a result, because of feasibility and efficiency concerns, we will derive analytically a generic close-form distribution expression for the PL between a centrally excited hexagon-based cell and a mobile device.

## 2. HEXAGON-BASED RNG

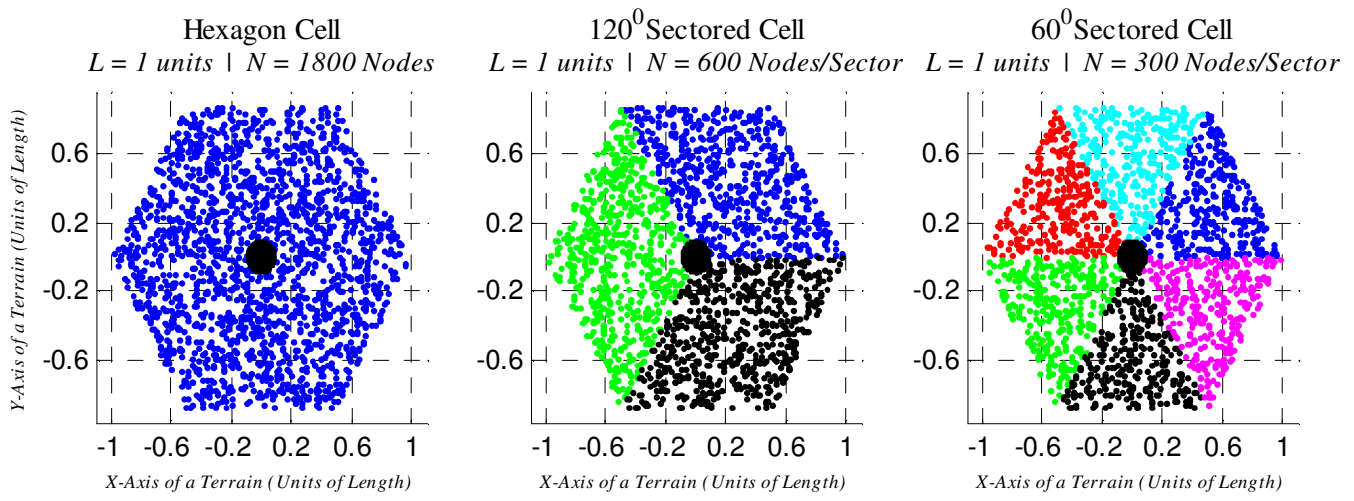

**Figure 1. Stochastic scattering for a hexagon cell.**

In order to drop random terminals inside a hexagon, as in Fig 1, we must first start by hypothesizing the joint spatial distribution of nodes. For simplicity and also since no spatial bias is noted inside a cell, we may assume a homogenous distribution within the featured area. This means that $f_{XY}(x,y) = 2/\left(3\sqrt{3}L^2\right)$, where $L$ represents the side length of a hexagon and $(x,y)$ is the spatial terrain coordinates of a node within the cell. Next, using stochastic methods, we obtain the marginal PDF along the $x$-axis, and from it with the help of the joint PDF we determine the marginal density $f_{Y|X=x_0}(y)$ for the $y$-axis. Furthermore, we find the Cumulative Distribution Function (CDF) for the $x$-component. Then, we obtain the inverse CDF as a function of some parameter "$u$" representing a sample from $U(0,1)$; where $U(a,b) = 1/(b-a)$ is a uniform distribution for $x \in (a,b)$. Using these findings, we could now apply the inverse transformation technique [5] to generate random samples for "$x$". Also, considering the $\{(b-a)u + a\}$ transformation assists in getting the "$y$" counterpart. Moreover, to ensure less interference and more users, 60° or 120° antenna sectoring can be applied [2]. As a consequence, proceeding in a similar fashion, we then obtain exact expressions for nodes dropping inside a triangle and a rhombus. An overall summary of the results is shown in Table 1.

**Table 1. Exact stochastic expressions for mobile dispersion**

| ***CELL SHAPES*** | ***RNG – $x_0$*** | ***RNG – $y_0$*** |
|---|---|---|
| Triangle 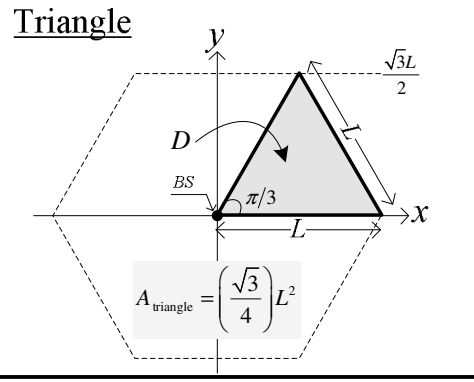  | $x_0 = \begin{cases} \sqrt{\frac{u_0}{2}}L & 0 < u_0 \le \frac{1}{2} \\ L\left(1-\sqrt{\frac{(1-u_0)}{2}}\right) & \frac{1}{2} \le u_0 < 1 \end{cases}$ | $f_{Y\|X=x_0}(y) = \begin{cases} U\left(0;\sqrt{3}x_0\right) & 0 < x_0 \le \frac{L}{2} \\ U\left(0;\sqrt{3}(L-x_0)\right) & \frac{L}{2} \le x_0 < L \end{cases}$ |
| Rhombus 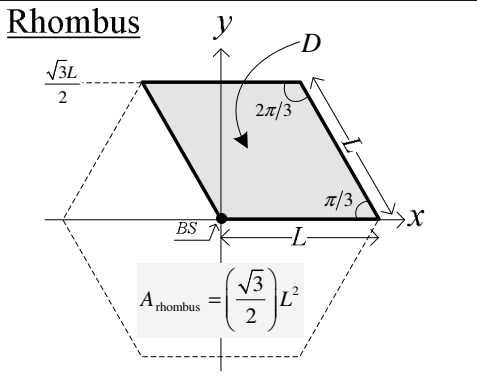  | $x_0 = \begin{cases} \frac{L}{2}\left(2\sqrt{u_0}-1\right) & 0 < u_0 \le \frac{1}{4} \\ L\left(u_0-\frac{1}{4}\right) & \frac{1}{4} \le u_0 \le \frac{3}{4} \\ L\left(1-\sqrt{1-u_0}\right) & \frac{3}{4} \le u_0 < 1 \end{cases}$ | $f_{Y\|X=x_0}(y) = \begin{cases} U\left(-\sqrt{3}x_0;\frac{\sqrt{3}L}{2}\right) & \frac{-L}{2} < x_0 \le 0 \\ U\left(0;\frac{\sqrt{3}L}{2}\right) & 0 \le x_0 \le \frac{L}{2} \\ U\left(0;\sqrt{3}(L-x_0)\right) & \frac{L}{2} \le x_0 < L \end{cases}$ |
| Hexagon 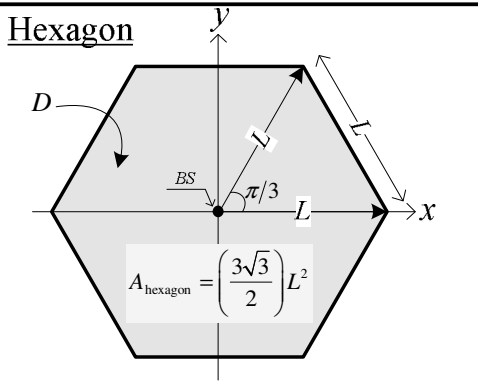  | $x_0 = \begin{cases} L\left(\sqrt{\frac{3u_0}{2}}-1\right) & 0 < u_0 \le \frac{1}{6} \\ \frac{3L}{4}(2u_0-1) & \frac{1}{6} \le u_0 \le \frac{5}{6} \\ L\left(1-\sqrt{\frac{3(1-u_0)}{2}}\right) & \frac{5}{6} \le u_0 < 1 \end{cases}$ | $f_{Y\|X=x_0}(y) = \begin{cases} U\left(\frac{-\sqrt{3}L}{2};\frac{\sqrt{3}L}{2}\right) & \|x_0\| \le \frac{L}{2} \\ U\left(-\sqrt{3}(L-\|x_0\|);\sqrt{3}(L-\|x_0\|)\right) & \frac{L}{2} \le \|x_0\| < L \end{cases}$ |

If we assume a constant density of say ≈ 6928 random Nodes/Units$^2$ for all simulated cases, then as evident from Fig. 2 samples from a histogram properly overlap theoretical derivations. This further justifies the analysis in addition to the spatial interpretation of Fig. 1.

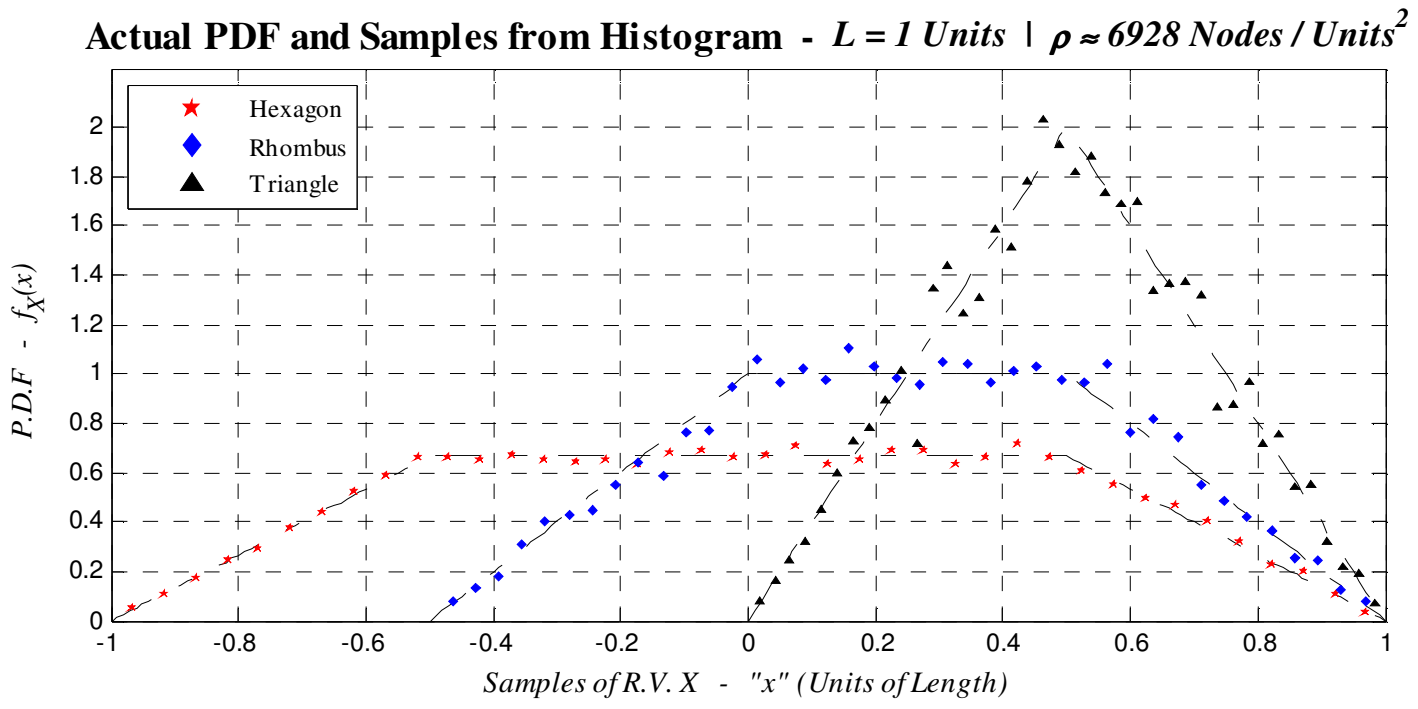

**Figure 2. Theoretical and experimental PDFs along the x-axis.**

A more detailed treatment on hexagon-based stochastic simulation is available in [6], which supports a complex cellular network with varying position, capacity, size, and users' density.

# 3. PATH-LOSS DISTRIBUTION MODEL

The spatial location of nodes will directly affect wireless factors among terminals, most notably channel corruptions and losses such as: path, shadowing and fading with probing accuracy of $\Delta \approx$ *1000λ*, *40λ*, and $\lambda$ meters for each case respectively; where $\lambda$ is the wavelength of the carrier frequency. If we monitor the channel attenuation for large and medium scale intervals and ignore the small scale, then it becomes natural to only look at the combination of PL and Shadowing. There are various models for the PL, though the most generic and widely used form has distance dependency as given by:

$$PL(r)_{dB} = \overline{PL(r)}_{dB} + \Psi = \overline{PL(r_0)}_{dB} + 10n\log_{10}\left(\frac{r}{r_0}\right) + \Psi \qquad (1)$$

where $\overline{PL(r)}_{dB}$ is the average PL in decibels (dB), "$\Psi$" is a Random Variable (RV) measured in dB representing the effect of shadowing with a log-normal distribution, i.e. $N(0,\sigma_\Psi^2)$, such that $N(m,\sigma^2)$ is a general Gaussian curve, and "$\sigma_\Psi$" is the standard deviation for shadowing also in dB. Further, "$n$" is the PL exponent which depends on the propagation environment such as the existence of Line of Sight [LOS] or otherwise. Also, "$r_0$" is referred to as the close-in distance measured in meters, and $r \ge r_0$ is the separation distance between a transmitter and a receiver. It is worth adding that the average PL at the close-in distance can be obtained empirically or for the simplest case, through the use of the Friis free space model. For ease of mathematical manipulations, we may represent (1) by mapping it to:

$$L_P = \overbrace{\alpha + \beta\log_{10}\left(\frac{r}{r_0}\right)}^{Path-Loss} + \overbrace{\Psi}^{Shadowing} = W + \Psi \qquad (2)$$

First, we need to find the density of "$r$" which is the distance between an in-cell mobile and the corresponding BS. Since a hexagon is really made-up of six equilateral triangles, we could only focus our model for nodes spreading in a 60° sector. Hence, the joint PDF of nodes spatial position in Cartesian coordinates becomes $f_{XY}(x,y) = 4/(\sqrt{3}L^2)$. Now, if we transform this distribution to polar notation we get:

$$f_{R\theta}(r,\theta) = \left[f_{XY}(x,y)\right]_{\substack{x=r\cos(\theta)\\ y=r\sin(\theta)}} \left|J(r,\theta)\right| = \frac{4r}{\sqrt{3}L^2} \qquad (3)$$

where $J(r,\theta)$ is a 2D Jacobian matrix. Next, using the law of sines with the help of Fig. 3, we obtain:

$$r = \frac{\sqrt{3}L}{2\sin(2\pi/3-\theta)} \qquad 0 \le \theta \le \pi/3 \qquad (4)$$

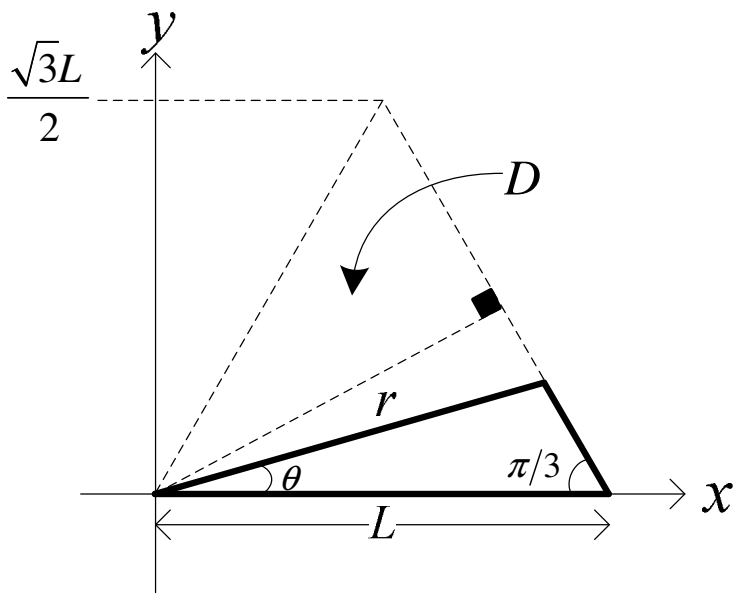

**Figure 3. Hexagon geometry for obtaining the radius.**

At this level, after utilizing (3) and (4) we find the following marginal PDF:

$$f_R(r) = \int_{\theta(r)\in D} f_{R\theta}(r,\theta)\,d\theta = \begin{cases} \dfrac{4\pi r}{3\sqrt{3}L^2} & 0 \le r \le \dfrac{\sqrt{3}L}{2} \\ \dfrac{8r}{\sqrt{3}L^2}\left\{\arcsin\left(\dfrac{\sqrt{3}L}{2r}\right) - \dfrac{\pi}{3}\right\} & \dfrac{\sqrt{3}L}{2} \le r \le L \end{cases} \tag{5}$$

If we now go back to (2) and focus on obtaining a distribution for the PL component only, while remembering that "$\alpha$" and "$\beta$" are deterministic scalars, we get:

$$f_W(w) = \frac{r_0 \ln(10)}{\beta} 10^{(w-\alpha)/\beta} \left[ f_R(r) \right]_{r = r_0 10^{(w-\alpha)/\beta}} = \begin{cases} \dfrac{4\pi r_0^2 \ln(10)}{3\sqrt{3}L^2\beta} 10^{2(w-\alpha)/\beta} \\ -\infty < w \le \underbrace{\alpha + \beta\log_{10}\left(\sqrt{3}L/2r_0\right)}_{\equiv a} \\ \dfrac{8r_0^2\ln(10)}{\sqrt{3}L^2\beta} 10^{2(w-\alpha)/\beta}\left\{\arcsin\left(\dfrac{\sqrt{3}L}{2r_0 10^{(w-\alpha)/\beta}}\right) - \dfrac{\pi}{3}\right\} \\ \underbrace{\alpha + \beta\log_{10}\left(\sqrt{3}L/2r_0\right)}_{\equiv a} \le w \le \underbrace{\alpha + \beta\log_{10}\left(L/r_0\right)}_{\equiv b} \end{cases} \tag{6}$$

Pursuing this further, it should stressed that both RVs "$W$" and "$\Psi$" for PL and shadowing are statistically independent, i.e. $\mathrm{E}[W\Psi] = \mathrm{E}[W]\mathrm{E}[\Psi]$. Therefore, to obtain the PDF of "$L_P$" we would need to convolve the density of both terms as shown by:

$$f_{L_P}(l) = f_W(l) * f_\Psi(l) = f_\Psi(l) * f_W(l) = \int_{-\infty}^{\infty} f_\Psi(\tau) f_W(l-\tau)\,d\tau \qquad l \in \mathbb{R} \tag{7}$$

where the components inside the integral as a function of a dummy variable "$\tau$" are given by (8) and (9). Note that the constants "$a$" and "$b$" used in (9) were defined within the limits of the density in (6).

$$f_\Psi(\tau) = N\left(0, \sigma_\Psi^2\right) = \frac{1}{\sqrt{2\pi}\sigma_\Psi} e^{-\tau^2/2\sigma_\Psi^2} \qquad \tau \in \mathbb{R} \tag{8}$$

$$f_W(l-\tau) = \begin{cases} 0 \quad \tau < l - b \\ \left\{\dfrac{8r_0^2\ln(10)}{\sqrt{3}L^2\beta}\right\} 10^{2(l-\tau-\alpha)/\beta} \times \arcsin\left(\dfrac{\sqrt{3}L}{2r_0 10^{(l-\tau-\alpha)/\beta}}\right) \\ l - b \le \tau \le l - a \\ \left\{\dfrac{-8\pi r_0^2 \ln(10)}{3\sqrt{3}L^2\beta}\right\} 10^{2(l-\tau-\alpha)/\beta} \\ l - b \le \tau \le l - a \\ \left\{\dfrac{4\pi r_0^2\ln(10)}{3\sqrt{3}L^2\beta}\right\} 10^{2(l-\tau-\alpha)/\beta} \\ l - a \le \tau < \infty \end{cases} \tag{9}$$

Also, it is important to realize that after several mathematical manipulations and labor the exponential parts (i.e. "$e$" and "$10$") could be modified to have the "$\tau$" entities combined together:

$$10^{2(l-\tau-\alpha)/\beta} e^{-\tau^2/2\sigma_\Psi^2} = e^{2\ln(10)\left\{\ln(10)\sigma_\Psi^2 + \beta(l-\alpha)\right\}/\beta^2} \times e^{-\left(\tau + 2\ln(10)\sigma_\Psi^2/\beta\right)^2/2\sigma_\Psi^2} \tag{10}$$

Finally, after performing the convolution, we obtain the *exact close-form* PL density as shown below, where $l \in \mathbb{R}$:

$$f_{L_P}(l) = \left\{\frac{4r_0^2\ln(10)}{\sqrt{3}L^2\beta}\right\} 10^{2\left\{\ln(10)\sigma_\Psi^2 + \beta(l-\alpha)\right\}/\beta^2} \times \left\{ \pi Q(\kappa_2) - \frac{2\pi}{3} Q(\kappa_1) + \frac{2}{\sqrt{\pi}} \int_{\kappa_1/\sqrt{2}}^{\kappa_2/\sqrt{2}} e^{-v^2} \arcsin\left(\frac{\sqrt{3}L}{r_0 10^{\left(\mu - \sqrt{2}\sigma_\Psi v\right)/\beta}}\right) dv \right\}$$

- $\mu = l - \alpha + 2\ln(10)\sigma_\Psi^2/\beta$
- $\kappa_1 = \left\{\mu - \beta\log_{10}\left(L/r_0\right)\right\}/\sigma_\Psi$
- $\kappa_2 = \left\{\mu - \beta\log_{10}\left(\sqrt{3}L/2r_0\right)\right\}/\sigma_\Psi$

(11)

In equation (11), the Q-function is a variation of the Error Function (ERF) or the Complementary Error Function (ERFC) as seen here:

$$Q(x)=\int_{x}^{\infty}N(0,1)\,du=\frac{1}{\sqrt{2\pi}}\int_{x}^{\infty}e^{-u^2/2}du$$
$$=\frac{erfc\left(x/\sqrt{2}\right)}{2}=\frac{\left\{1-erf\left(x/\sqrt{2}\right)\right\}}{2} \qquad (12)$$

An infinite series equivalent of $Q(x)$ can be found in say [7]. As for the integral part of (11), it can be evaluated using any numerical integration method such as Simpson's or trapezoidal rule. Though, we were also capable to find a close expression for this integration as detailed in the Appendix section of this paper.

Before proceeding to simulation, we should mention that the close-form model for the generic PL derived here will equally be the same for sectored or non-sectored cells, because after all the fundamental contour of Fig. 3 is repeated for a rhombus and a hexagon shape. And, the PL only depends on the separation between a BS and an MS, represented by the radius "$r$", and does not take into account the fundamental sector's angle of rotation about the origin.

# 4. SIMULATION RESULTS

For simulation, we will utilize parameters from the MBWA standard [8]. The values used for different channel environments are shown in Table 2, where the carrier frequency assumed is 1.9 GHz. Also, in the table, the close-in distance is not explicitly given, and is in fact absorbed by "$\alpha$" of the PL, in other words:

$$W=\alpha+\beta\log_{10}\left(\frac{r}{r_0}\right)=\left\{\alpha-\beta\log_{10}\left(r_0\right)\right\}+\beta\log_{10}\left(r\right) \qquad (13)$$
$$=\alpha'+\beta\log_{10}\left(r\right)$$

**Table 2. IEEE 802.20 channel models**

| **Channel Environment** | *Suburban Macrocell* | *Urban Macrocell* | *Urban Microcell* | |
|---|---|---|---|---|
| **Cell Radius** [km] | *0.6 ≤ L ≤ 3.5* | *0.6 ≤ L ≤ 3.5* | *0.2 ≤ L ≤ 0.3* | |
| **Propagation Model** | *COST-231 Hata-Model* | *COST-231 Hata-Model* | *COST-231 Walfish-Ikegami* | |
| **Standard Deviation for Shadowing** [dB] | *10* | *10* | *NLOS* | *LOS* |
| | | | *10* | *4* |
| **Path-Loss** [dB] | *ά =31.5 β =35* | *ά =34.5 β =35* | *ά =34.53 β =38* | *ά =30.18 β =26* |
| **Supported Distances** [m] | *r₀ = 35 ≤ r ≤ L* | *r₀ = 35 ≤ r ≤ L* | *r₀ = 20 ≤ r ≤ L* | |
| **Mobility** [km/hr] | *0 → 250* | *0 → 250* | *0 → 120* | |

The results, based on random dropping of 10,000 terminals for each case, under different channel parameters, are shown in Fig. 4. As it can be observed, the values properly match the theoretical close-form expression derived in this treatment.

# 5. CONCLUSION

Analysis based on a hexagon cell is always preferred as oppose to a circular one because of tessellations. However, to the best of our knowledge, though because of simplicity they have been shown for the circular case [3] [4] [9], no hexagon-based models in literature are available for random nodes scattering or PL density. Therefore, in this paper, we for the first time obtained exact close-form stochastic results for both of these objectives. We also verified the analytical derivation through simulation and as expect both scattering and PL distribution match the theory. Moreover, because the models were deliberately derived with generic parameters, they are hence practical and can be applied for any cellular technology during the design phase by system engineers. Nonetheless, to demonstrate the analysis, we have based our simulation using specifications from the new MBWA IEEE 802.20 protocol.

# 6. APPENDIX

Equation (11) has an integration with an exponent and an arcsine. As mentioned earlier, numerical methods could be used to solve this. However, after careful manipulations the integration turns out to exit in close-form, and we thought it is worth mentioning it here. Note that we will not go through the entire derivation because it is very long and will defeat the purpose of the paper. We will suffice by mentioning only the major highlights and steps needed to converge to a solvable equality.

Since, the integral part is of the form:

$$I=\int_{x_1}^{x_2}e^{-x^2}\arcsin\left(\frac{k}{10^{\{a+bx\}}}\right)dx \qquad (A.1)$$

we start by utilizing calculus's integration by parts, where "$u$" is assigned to the exponent part and "$dv$" to arcsine. The derivative of "$u$" is obtained in a straightforward manner. As for the integration of "$dv$", we substitute a variable for the "$a+bx$" term. After doing this, the integral becomes of the form $\arcsin(y)/y$. Now, we could not solve this without relying on infinite Taylor series. A sequence expression for the arcsine is available in any handbook on mathematics, such as [10]. Then, we divide the arcsine series by "$y$" and take the integral to get "$v$". At this level, we notice that the second half of the integration by parts will have two components of the form:

$$\int v\,du=\int xe^{-x^2}10^{-\eta x}dx \qquad (A.2)$$

Hence, we combine the exponential terms, similar to what we did in (10), and then perform yet another integration by parts, followed by direct substitution. After several steps of manipulations, we finally converge to the expression shown here:

$$I=\left\{\frac{k}{\gamma 10^{a}}\right\}\left\{\begin{array}{l}\sum_{n=0}^{\infty}C_n e^{\xi^2}\left[e^{-\{x+\xi\}^2}-2\sqrt{\pi}\xi Q\left(\sqrt{2}\left(x+\xi\right)\right)\right]\\ -e^{-x^2}\left[10^{-bx}+\sum_{n=1}^{\infty}C_n 10^{-b(2n+1)x}\right]\end{array}\right\}_{x_1}^{x_2}$$

- $\gamma=b\ln(10)$
- $\xi=\gamma(2n+1)/2$
- $C_n=\begin{cases}1 & n=0\\ \dfrac{(2n-1)!\,\left(k10^{-a}\right)^{2n}}{2^{2n-1}(n-1)!\,n!\,(2n+1)^2} & n=1,2,\ldots,\infty\end{cases}$ ■

(A.3)

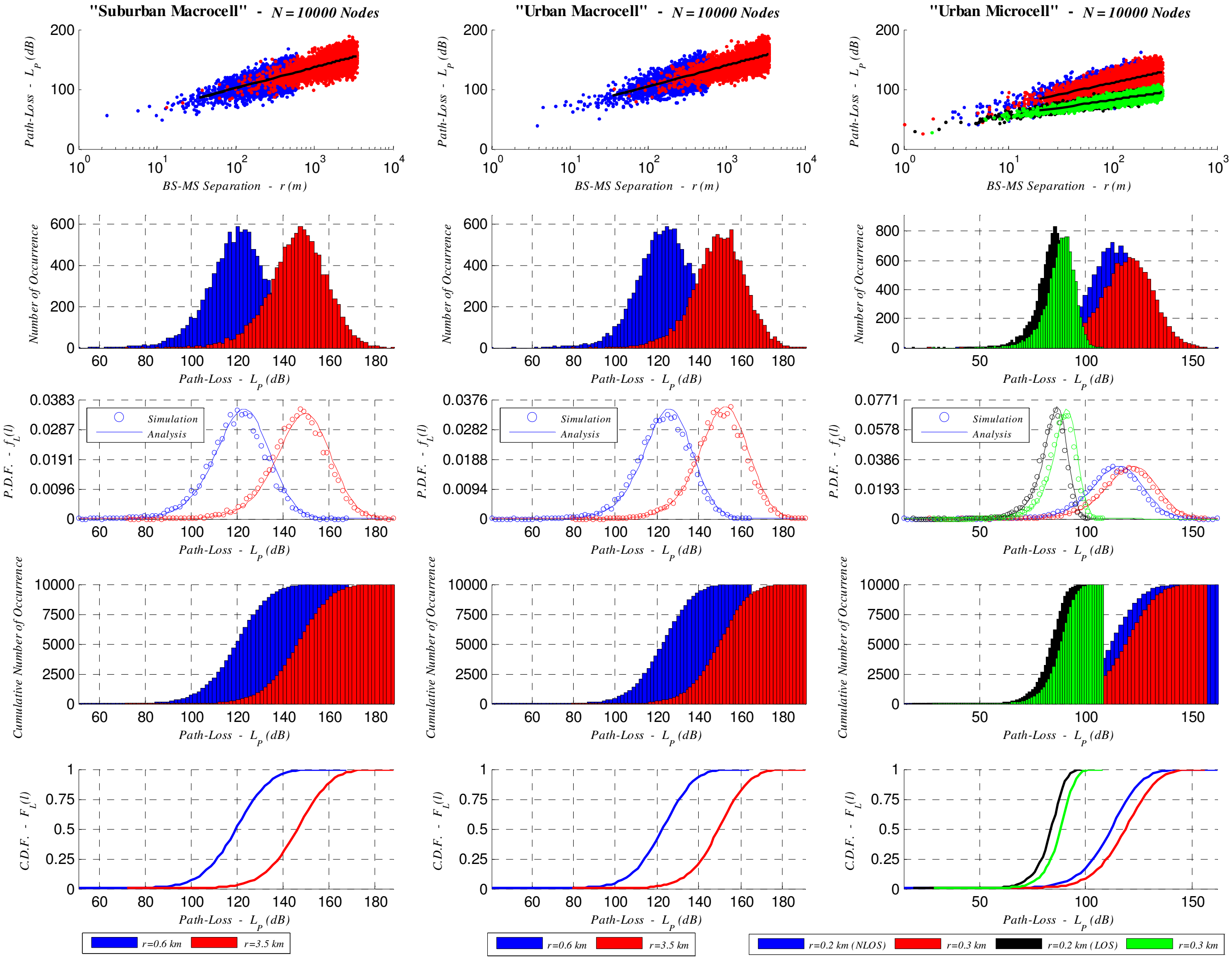

**Figure 4. PL Simulation under different IEEE 802.20 channel environments.**